\documentclass[a4paper,11pt]{article}
\usepackage{geometry}
\usepackage[american]{babel}
\usepackage{amsmath,amsfonts,amsthm,amssymb}
\usepackage{bbm}
\usepackage[utf8]{inputenc}
\usepackage{csvsimple}
\usepackage{fontenc}
\usepackage{graphicx}
\usepackage{epstopdf}
\usepackage{caption}
\usepackage{exceltex}
\usepackage{ulem}
\usepackage{color}
\usepackage{natbib}
\usepackage{nicefrac}
\usepackage{booktabs}
\usepackage{enumitem}
\usepackage{lscape}
\usepackage{setspace}
\setitemize{leftmargin=*}
\usepackage{hyperref}
\usepackage{eurosym}
\usepackage{multirow}

\date{}
\author{
Malte Jahn\thanks{Department of Mathematics and Statistics, Helmut Schmidt University, 22043 Hamburg, Germany. E-Mail: \href{mailto:jahnma@hsu-hh.de}{\nolinkurl{jahnma@hsu-hh.de}}. ORCID:  \href{https://orcid.org/0000-0002-7165-0428}{0000-0002-7165-0428}.}
}

\title{\textbf{Artificial neural networks and time series of counts: A class of nonlinear INGARCH models}}

\begin{document}
\maketitle
\bigskip
\bigskip

\vspace{2cm}
\noindent
\begin{center}
\textbf{Abstract}\\
Time series of counts are frequently analyzed using generalized integer-valued autoregressive models with conditional heteroskedasticity (INGARCH). These models employ response functions to map a vector of past observations and past conditional expectations to the conditional expectation of the present observation. In this paper, it is shown how INGARCH models can be combined with artificial neural network (ANN) response functions to obtain a class of nonlinear INGARCH models. The ANN framework allows for the interpretation of many existing INGARCH models as a degenerate version of a corresponding neural model. Details on maximum likelihood estimation, marginal effects and confidence intervals are given. The empirical analysis of time series of bounded and unbounded counts reveals that the neural INGARCH models are able to outperform reasonable degenerate competitor models in terms of the information loss.
\end{center}

\bigskip
\bigskip

\newpage

\section{Introduction}
Artificial neural networks (ANN) are state-of-the-art tools for data analysis. They are often applied in big data contexts but also for classical problems such as non-linear regression and time series analysis \citep[e.g.][]{Tkacz01,KockTeraesvirta14, Jahn20}, which is the relevant setting of this paper. There are different types of ANN but the main reason to use them is almost always their universal approximation property \citep[cf.][p.\ 616]{KockTeraesvirta14}. In most applications, the ANN function is fitted (``trained'') to data by minimizing the mean squared error. In regression analysis, this (nonlinear) least squares approach is very appropriate in case of a normal error distribution, but less appropriate for other distributions, in particular skewed distributions. In this paper, time series of counts are considered and the corresponding count distributions are often not well approximated by a normal distribution. An important class of models for this type of data are integer-valued autoregressive models with conditional heteroskedasticity (INARCH). These models assume a conditional count distribution at every time step where the conditional mean depends on past time series values. Natural choices are the Poisson distribution (for unbounded counts) or the binomial distribution (for bounded counts). Since, for these count distributions, the conditional variance is directly related to the conditional mean parameters, the models exhibit conditional heteroskedasticity. INARCH models can be discussed in their more general version known as INGARCH \citep{ferland06, Fokianos09}. By including past conditional means as additional regressors, past conditional variances are included implicitly which makes the INGARCH model similar to the traditional GARCH approach for continuous distributions.\\
A frequent issue with exactly linear INGARCH models is that the conditional mean parameter may leave the support of the count distribution. One solution is to impose quite strict parameter constraints such as positive coefficients in exactly linear INGARCH models \citep[][eq.\ 1]{ferland06}. The downside is that these parameter constraints also limit the ability to account for negative autocorrelation. The alternative solution is to employ response functions which map values (i.e.\ conditional mean parameters) outside of the plausible range to the known domain while maintaining desirable properties, such as differentiability. Depending on the probability distribution, choices are a log-linear link leading to a multiplicative model \citep{fokianostjostheim11} or the logit link \citep{chen20}. We refer to a link function as the inverse response function (if exists), such that, e.g.\,, a log link corresponds to an exponential response and a logit link corresponds to a logistic response. Recent approaches are the softplus response for unbounded counts \citep{WeissZH22} and the soft-clipping response for bounded counts \citep{weissjahn21} which yield nearly linear models but may still account for negative autocorrelation. In this paper, nonlinear INGARCH models are considered where the response function corresponds to a single hidden layer feedforward artificial neural network (neural INGARCH model).\\
In the following section, neural network architectures are discussed and it is shown how the ANN concept of activation functions relates to the INGARCH concept of response functions and why many existing INGARCH models can be interpreted as a degenerate version of a corresponding neural INGARCH model. A short section considers model selection strategies via information criteria and diagnostic quantities derived from the (standardized) Pearson residuals. The first data example then considers the number of (countries experiencing) banking crises in a given year from 1800 to 2011. For the neural INGARCH model, a time variable allowing for a non-linear trend proves useful to represent a period between the 1940s and the 1960s when almost no banking crisis occurred. The remaining overdispersion can be modeled by replacing the conditional Poisson distribution by a generalized Poisson distribution. As a second data example for time series of bounded counts, the number of votes in favor of an adjustment of the interest rate by the members of the Monetary Policy Council (MPC) of the National Bank of Poland between 2002 and 2013 are considered. The previous results are confirmed, the neural INGARCH model also outperforms the corresponding ``degenerate'' conventional INGARCH model in terms of the information loss. The model can be further improved by including a discrete policy regime variable. The final section concludes.

\section{The neural INGARCH model}
An INGARCH(p,q) model is defined via the conditional mean equation
\begin{equation}\label{ingarch}
E[Y_t|\mathcal{F}_{t-1}]=\lambda_t=f(Y_{t-1},\dots,Y_{t-p},\lambda_{t-1},\dots,\lambda_{t-q}),
\end{equation}
for some response function $f$. The actual conditional distribution of $Y_t$ is an appropriate count distribution with mean $\lambda_t$. The right hand side depends on $p$ past values of the series and on $q$ past conditional means. A linear response function $f(x)=x$ would lead to the exactly linear INGARCH model \citep[][eq.\ 1]{ferland06}. In this paper, response functions from the class of artificial neural networks are considered. More precisely, it is assumed that $f$ is a single hidden layer feedforward network (SLFN). The mathematical formulation of the SLFN regression function is:
\begin{equation}\label{annreg}
f^{ANN}(w^0,w^1,x)=g_1\! \left(\sum_{h=1}^{H}w^1_{h}\cdot g_0\! \left(\sum_{j=1}^{J}w^0_{jh} x_j\right)\right).
\end{equation}
The SFLN function depends on a set of parameters $w^{1}_{h}$ and $w^{0}_{jh}$. The input vector $x$ consists of a constant and the respective lagged values, so $x=(1,Y_{t-1},\dots, Y_{t-p},\lambda_{t-1},\dots,\lambda_{t-q})$ and its number of elements is $J=p+q+1$. The index $h$ ($1 \leq h \leq H$) refers to the so-called hidden neurons. Their number $H$ determines the complexity of the model. In practice, the optimal $H$ can be determined by information criteria, which will be discussed later. The functions $g_0$ and $g_1$ are activation functions which are required to be increasing and continuously differentiable. Since $g_0$ and $g_1$ are generally nonlinear, it is possible to give a basic interpretation of the corresponding neural INGARCH model: The conditional expectation of $Y_t$ is a nonlinear combination of nonlinear combinations of $Y_{t-1},\dots,Y_{t-p},\lambda_{t-1},\dots,\lambda_{t-q}$.\\

\begin{figure}[!h]
\center
\includegraphics[width=0.5\textwidth, trim={2cm 0.9cm 3cm 1cm},clip,page=3]{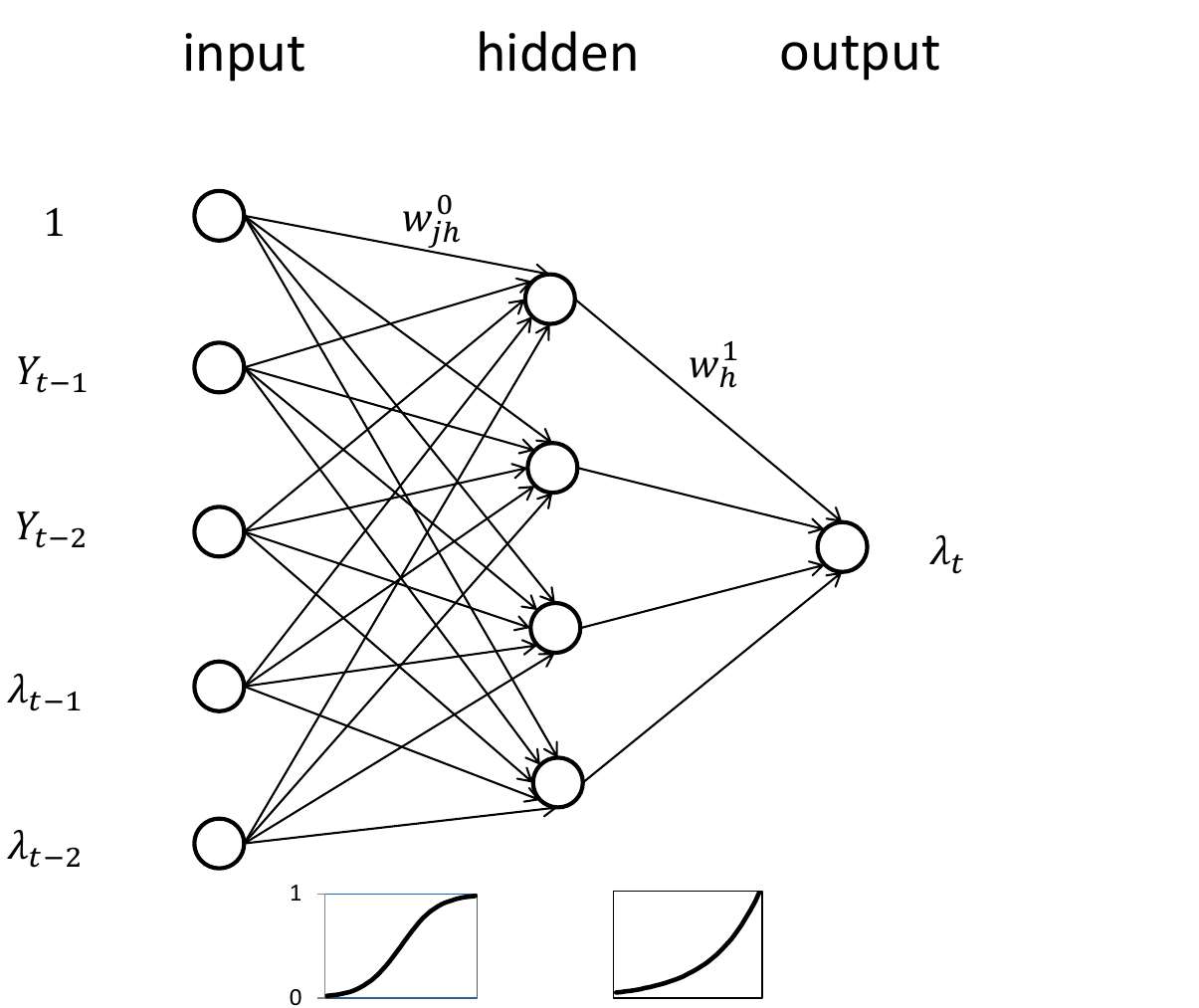}
\caption{Visual representation of a softplus Poisson neural INGARCH(2,2) model with $H=4$ hidden neurons}
\label{netvisual}
\end{figure}

Figure \ref{netvisual} shows a concrete specification of a neural INGARCH(2,2) model and helps explaining the concept of hidden neurons and activation functions. The SLFN consists of three layers: one input layer, one hidden layer and one output layer. The neurons (circles) in the input layer correspond to the regressor variables, the hidden layer is unobserved and a specification with 4 hidden neurons is chosen for illustration. The single neuron in the output layer corresponds to the conditional mean of the current value of the time series. Alternatively, instead of the output, an illustration based on the target value $Y_t$ could be used.  As indicated by the direction of the arrows, it is assumed that information is only passed in the forward direction from the input layer through the hidden layer to the output layer, which makes the network a single hidden layer ``feedforward'' network.\\

The interesting elements for this paper are the activation functions $g_0$ and $g_1$ which are sketched at the bottom of Figure \ref{netvisual}. The term ``activation'' refers to the name-giving biological interpretation: A neuron passes on information depending on its level of activation which is determined by the information in the previous layer. For the activation of the hidden layer, the logistic function is employed throughout the paper which introduces (desired) non-linearity into the model, so $g_0(x)=\frac{1}{1+\exp(-x)}$. This also implies that the values of the hidden neurons are always between 0 (completely inactive) and 1 (completely active). The activation of the output layer requires a deeper consideration and one has to look at the data to make an appropriate choice for $g_1$. The depicted function corresponds to the softplus function, so $g_1(x)=\ln(1+\exp(x))$. The softplus function can be seen as a smoothed version of the ReLU function. The image of the function are the positive real numbers, so it makes sense to use it in the case of unbounded counts. In general, the output activation function $g_1$ takes the same role as the response function in various INGARCH models \citep[e.g.][]{chen20, weissjahn21, WeissZH22}. Such ``conventional'' INGARCH models can be interpreted as degenerate neural INGARCH models where the hidden layer is missing and the output is directly activated from the inputs through the function $g_1$.  Conversely, when the starting point is a certain conventional INGARCH model with response function $g$, it is possible to construct a corresponding neural model by setting $g_1=g$ in many cases. One can then use statistical methods such as information criteria to check whether it is reasonable to ``upgrade'' that particular conventional model to a neural model by adding the hidden layer.\\

Assuming the generic conditional Poisson distribution for unbounded counts, the conditional log-likelihood function of an INGARCH model with response function $f$ and parameter vector $w$ is
\begin{equation}
\ell(w)=\sum_{t}Y_t \ln(f(w,x))-f(w,x)+c, 
\end{equation}
which has to be maximized w.r.t.\ $w$. For practical modeling, it shall be noted here that the numerical maximization speed can be significantly improved in the case of a purely autoregressive model ($q=0$) because then, the regressors are predetermined and can be collected as columns of a design matrix $X$ as usual for standard regression models. On the other hand, if $q>0$, the log-likelihood function (and the gradient) are defined recursively, implying that an optimization algorithm needs to loop through the individual time periods in each iteration. According to the chain rule, we have the following relationship:
\begin{equation}
\frac{\partial \ell}{\partial w}=\frac{\partial \ell}{\partial f}\frac{\partial f}{\partial w}.
\end{equation}

Regarding the gradient, the outer derivative is simply the derivative of the conditional Poisson log-likelihood with respect to the Poisson parameter $f(w,x)$: 
\begin{equation}\label{foc}
\frac{\partial \ell}{\partial f}=\sum_{t}\left(\frac{Y_t}{f(w,x)}-1\right).
\end{equation}

The first-order condition derived from equation \eqref{foc} implies that the relative error (ratio between true and fitted values) is the relevant error for the Poisson model. The partial derivatives $\frac{\partial f}{\partial w}$ can be computed by further applications of the chain rule. In case of the ANN response function $f=f^{ANN}$, this can be done efficiently by using the so-called backpropagation procedure. The only difference to the conventional backpropagation is that the error which is back-propagated is now the relative error instead of the plain error which arises from the first-order condition of the least squares approach. For an efficient calculation, it makes sense to use activation functions with simple derivatives. In the present example, we have for the softplus function $g_{1}'=g_0$ (i.e.\ the derivative of the softplus function is the logistic function) and for the logistic function $g_{0}'=g_0(1-g_0)$. \\

\section{Model selection and diagnostics}\label{modsec}
From the previous section, it became clear that the relevant quantities regarding model selection in the neural INGARCH model are the model order $(p,q)$ and the number of hidden neurons $H$. Likelihood-based information criteria such as Akaike's information criterion (AIC) or the Bayesian information criterion (BIC) can be used to determine the optimal $(p,q)$ and $H$. The proposed model selection strategy is to select $(p,q)$ first based on the corresponding ``degenerate'' conventional INGARCH model. This establishes a reasonable simple competitor for the neural model and the further analysis can be focused on the question whether to ``upgrade'' this particular conventional model to a neural model and which complexity $H$ is necessary. There are also many rules of thumb for restricting $H$ in advance. Obviously, the number of available time periods $T$ prevents us from using excessively large $H$. For example, one may require 10 times more observations than parameters, meaning $H\leq\frac{0.1 T}{J+1}$ and assuming that the number of regressors $J$ has already been fixed.\\
For model diagnostic, a simple and informative quantity are the (standardized) Pearson residuals:
\begin{equation}\label{pearson}
r_{t}=\frac{Y_t-E[Y_{t}|\mathcal{F}_{t-1}]}{\sqrt{Var[Y_t|\mathcal{F}_{t-1}]}}.
\end{equation}
The empirical Pearson residuals can be computed easily after estimation, even for the otherwise complex neural model because the formula only involves predicted conditional means and variances. According to the analysis in \citet{weiss20}, the Pearson residuals can help to detect (remaining) overdispersion and serial correlation. In a correctly specified model, the Pearson residuals should have mean zero, variance one and no serial correlation. \citet{weiss20} also discuss asymptotic results allowing for formal tests under certain assumptions. In this paper, mainly the variance of the Pearson residuals will be used as a rough indicator for whether the dispersion behavior is represented correctly by a given model.

\section{Modeling unbounded counts}\label{pinarch}
\begin{figure}[!h]
\center
\includegraphics[width=0.5\textwidth, trim={0cm 0.cm 0cm 0cm},clip]{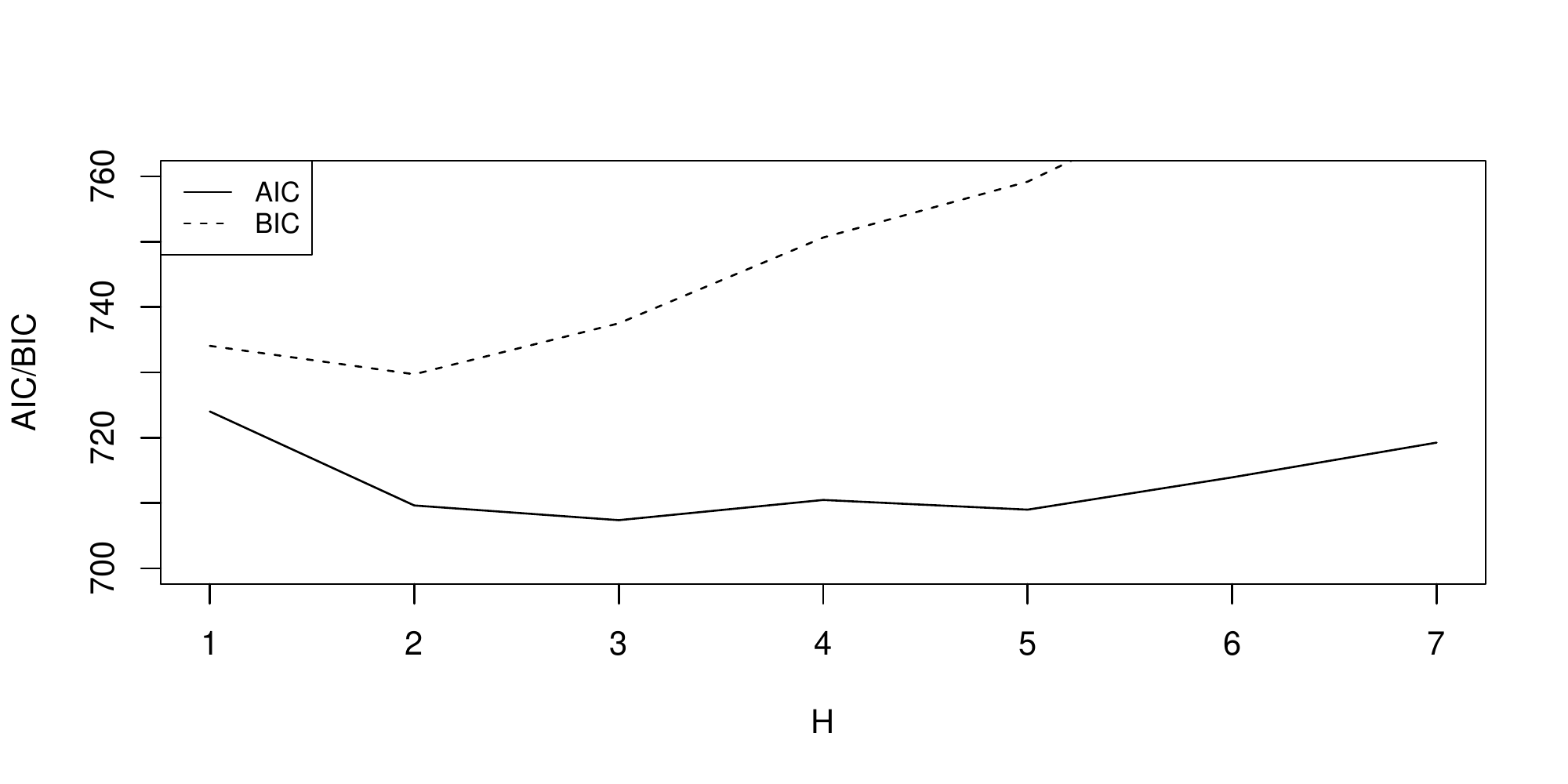}
\caption{Banking crises: Information loss of the neural Poisson INGARCH(1,0) for different $H$}
\label{modsecbank}
\end{figure}

As a first example, the annual number of countries experiencing a banking crisis among a group of 21 countries in the time period 1800 to 2010 (211 years) is considered. This data set was originally discussed by \citet{ReinhartRogoff09} but the main reference for this paper will be \citet{dungey20}. Following \citet{dungey20}, the data are treated as unbounded counts (although they are technically bounded). For the neural INGARCH model, this implies that the softplus function is chosen as the output activation function ($g_1(x)=\ln(1+\exp(x))$), in combination with a conditional Poisson distribution. The model definition becomes
\begin{equation}\label{poisson}
Y_t \sim Poi(\lambda_t) \mbox{ with } \lambda_t=f^{ANN}(Y_{t-1},\dots,Y_{t-p},\lambda_{t-1},\dots,\lambda_{t-q}).
\end{equation}\\
According to the model selection procedure proposed in the previous section, the first step is to determine the model order by looking at the corresponding conventional INGARCH model. Here, the relevant conventional model is the (nearly linear) softplus Poisson INGARCH model proposed by \citet{WeissZH22}. The regression function for this model is $f^{sp}(x,\beta)=g_1(x\beta)$. The analysis of the banking crisis data by \citet{dungey20} suggests that small model orders are sufficient. Indeed, Table \ref{order} (top) shows that the simplest model, the softplus Poisson INGARCH(1,0) is preferred over the larger alternatives INGARCH(2,0) and INGARCH(1,1) in terms of the information loss (AIC/BIC).\\
Once the model order is established, the final step in the specification of the neural INGARCH model is to determine the complexity $H$. The rule of thumb from section \ref{modsec} gives $H\leq\frac{0.1 T}{J+1}=\frac{21}{3}=7$. Figure \ref{modsecbank} shows that the AIC is minimal for $H=3$ and the BIC is minimal for $H=2$. Since the minimum of the BIC is a little bit more pronounced, $H=2$ is selected for now.

\begin{figure}[!h]
\center\normalsize
\includegraphics[width=\textwidth, trim=0cm 0cm 0cm 1.5cm, clip]{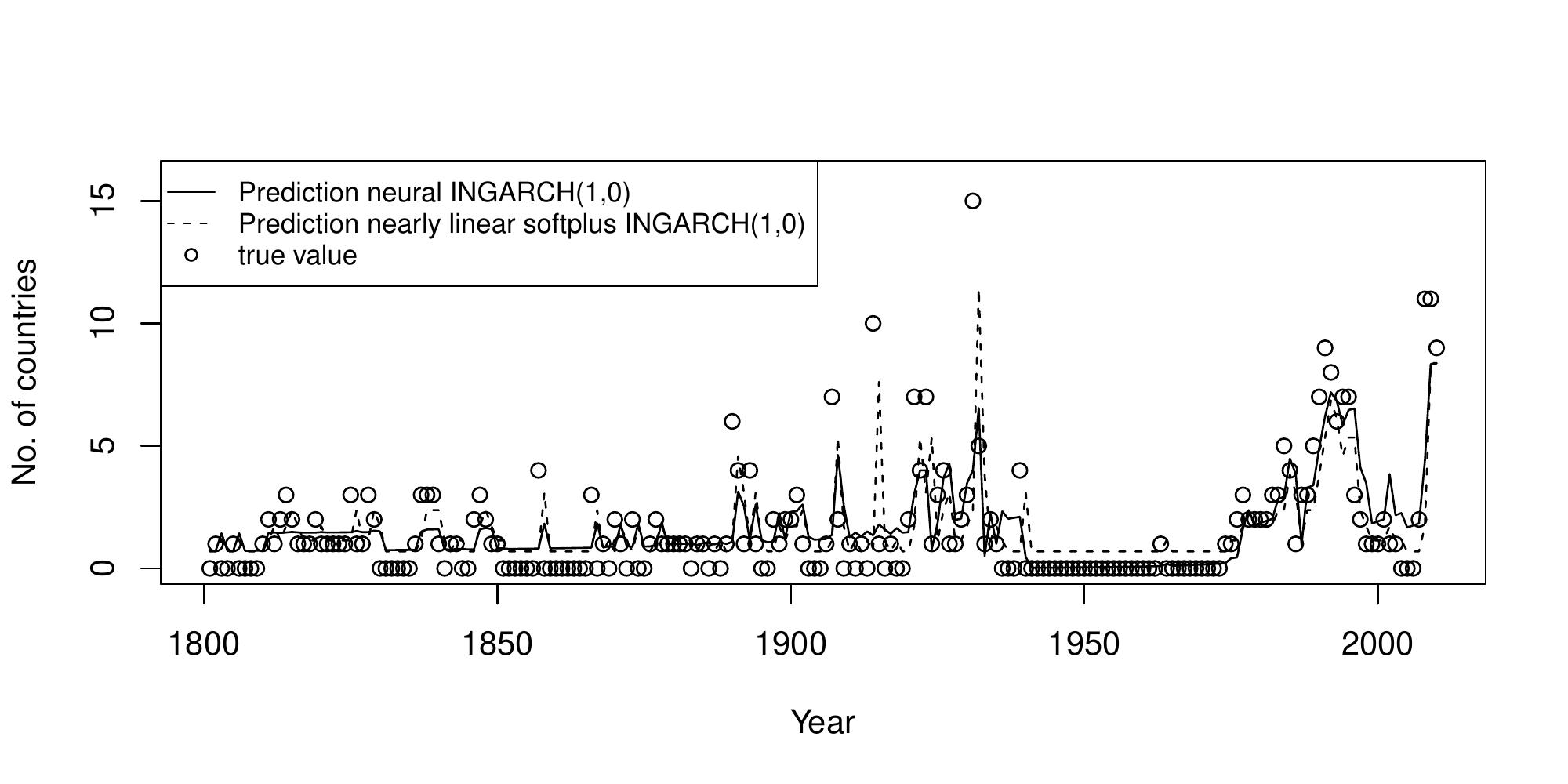}
\caption{Banking crises: Predicted conditional means by the neural INGARCH(1,0) model and the nearly linear softplus INGARCH(1,0) model}
\label{prediction}
\end{figure}

The precise AIC and BIC value for the neural Poisson INGARCH(1,0) with $H=2$ hidden neurons is displayed in Table \ref{order}. In comparison with the conventional competitor model, the neural model has lower AIC but higher BIC, so there is no clear evidence (yet) whether the higher likelihood is worth the additional parameters. Regarding interpretation, the estimated parameter values $\hat{w}$ are of minor interest because they have no direct meaning. Therefore, the initial step in the interpretation of the neural INGARCH model is to compare its predictions to the corresponding conventional model. Figure \ref{prediction} shows the prediction for the two models, along with the true values. The biggest challenge is to capture the long period (almost) without any banking crises between the 1940s and the 1960s when the Bretton-Woods (BW) system of fixed exchange rates was in place \citep[cf.][p.\ 2, bottom]{dungey20}.

\begin{figure}[!ht]
\center\footnotesize
\begin{tabular}{c@{\qquad}c}
\includegraphics[trim=0cm 0cm 0cm 0cm, clip,scale=0.45]{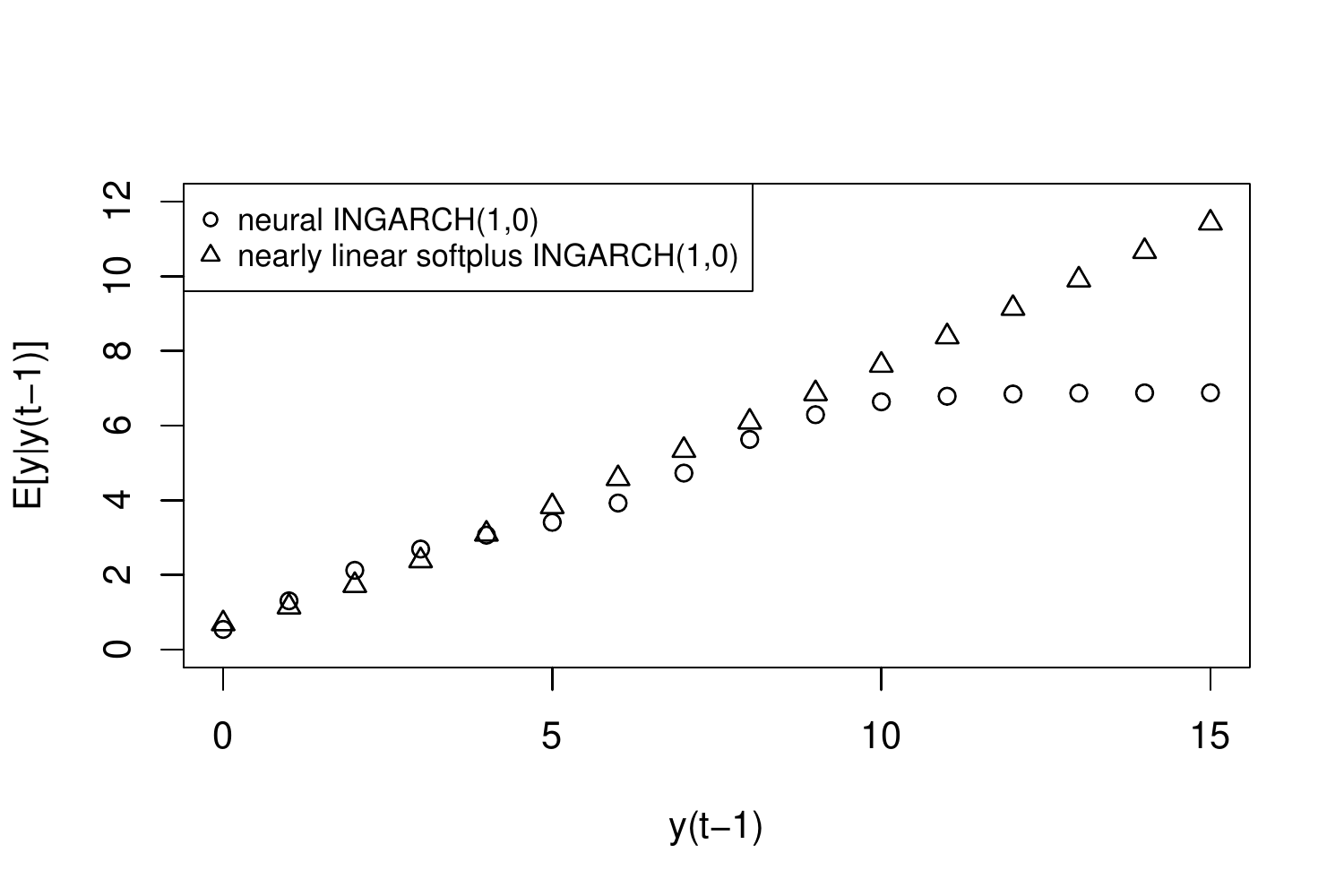} & \includegraphics[trim=0cm 0cm 0cm 0cm, clip,scale=0.45]{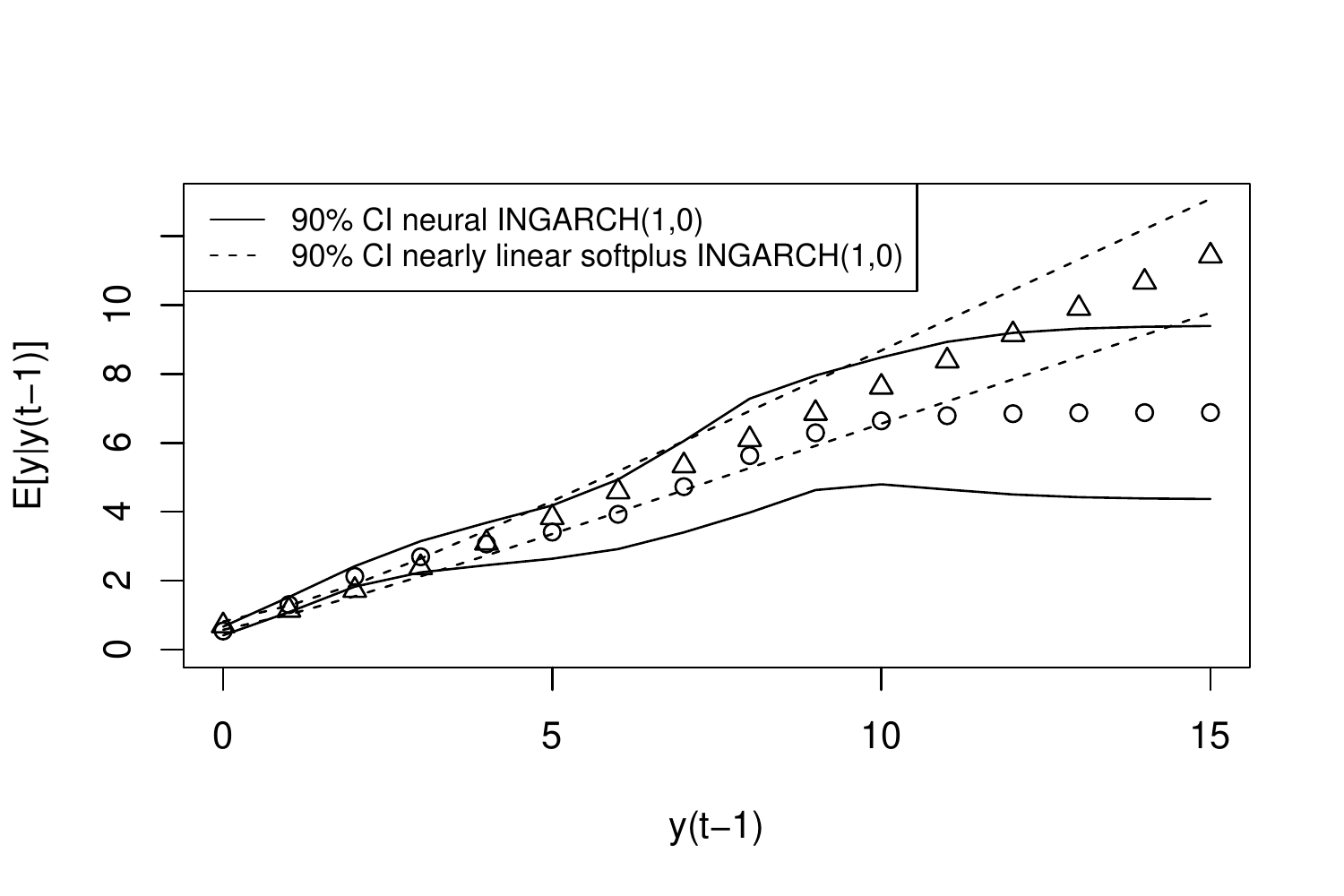} \\
(a)&(b)\\
\end{tabular}
\caption{Banking crises: Marginal effect of $y_{t-1}$ for the neural INGARCH(1,0) model and the nearly linear softplus INGARCH(1,0) model (a) and corresponding confidence intervals (b)}
\label{margbank}
\end{figure}

The second step of the analysis is concerned with marginal effects. As usual for non-linear regression models, there are different options to calculate marginal effects and it is usually advocated to consider marginal effects ``at the mean''. For the present model, there is only one variable which facilitates the illustration and interpretation. Figure \ref{margbank}~(a) shows plots of the predicted conditional mean of $y_t$ given different values of $y_{t-1}$. It becomes clear that, in the neural model, the marginal effect (i.e.\ the slope of the ``curve'') is not limited to a constant as in the (nearly) linear model. Although the differences between the two models are not very pronounced, the neural model indicates, for example, that the marginal effect of $y_{t-1}$ becomes very small for $y_{t-1}\geq 10$ which means that the neural model is more likely to return to a lower number of banking crises immediately after a year with a large number of banking crises. Regarding the ability to model the BW period, the probability of staying in the zero-count state is the relevant quantity and it can be concluded visually from both Figure \ref{prediction} and Figure \ref{margbank} that this probability is slightly higher for the neural model. Since a conditional Poisson distribution is assumed, this zero probability can be calculated easily. We have $p_{0|0}^{sp}=\exp(-g_1(\hat{\beta}_0))=0.50$ and $p_{0|0}^{ANN}=\exp(-f^{ANN}(\hat{w},x=(1,0))=0.58$.\\

For further inference, one might also want to calculate confidence intervals. Since the parameters have no direct interpretation, sensible confidence intervals should refer to the conditional mean $E[y|x]$. Corresponding predicted conditional means have been considered in the previous step of the analysis (Figure \ref{margbank}~(a)). Now, the focus is on the conditional variance of $f(x,\hat{w})$. For non-linear regression models, we generally have $Var(f(x,\hat{w}))\approx \nabla f(x)^T \mathcal{H}^{-1} \nabla f(x)$. The notation is such that $\hat{w}$ is the estimated parameter vector, $\nabla f(x)$ is the gradient of $f$ (w.r.t.\ $w$) evaluated at $x$ and $\mathcal{H}$ is the Hessian matrix of the model at $w=\hat{w}$. 
As noted earlier, the gradient $\nabla f(x)$ can be calculated efficiently in the neural INGARCH model due to the simple derivatives of the involved activation functions and the backpropagation method. Depending on the employed algorithm, the Hessian matrix $\mathcal{H}$ may be computed as part of the optimization procedure or afterwards. Further information on confidence intervals for ANN regression models in the usual least squares context can be found in \citet{RivalsPersonnaz00}.\\
Concerning the nearly linear softplus model with $f^{sp}(x,\beta)=g_1(x\beta)$, we have $Var(f^{sp}(x,\hat{\beta}))\approx \nabla f^{sp}(x) \mathcal{H}_{sp}^{-1} \nabla f^{sp}(x)$. For the softplus response function, the gradient (w.r.t.\ $\beta$) is $\nabla f^{sp}(x)=\frac{-x}{1+\exp(-x)}$. The denominator is almost unity for arguments $x\geq 2$. This shows that the softplus model behaves very similarly to an exactly linear Poisson INGARCH model when the time series is further away from zero.\\
Figure \ref{margbank}~(b) displays approximate 90\% pointwise confidence intervals for the conditional means for both the neural and the softplus model. The highest confidence is generally around the mean of the process which is just 1.55 in the present example. Banking crises affecting many countries at the same time are quite rare (cf.\ Figure \ref{prediction}) and therefore, the confidence is lower in that region.

\begin{figure}[!ht]
\center\footnotesize
\begin{tabular}{c@{\qquad}c}
\includegraphics[trim=0cm 0.5cm 0cm 1cm, clip,scale=0.45]{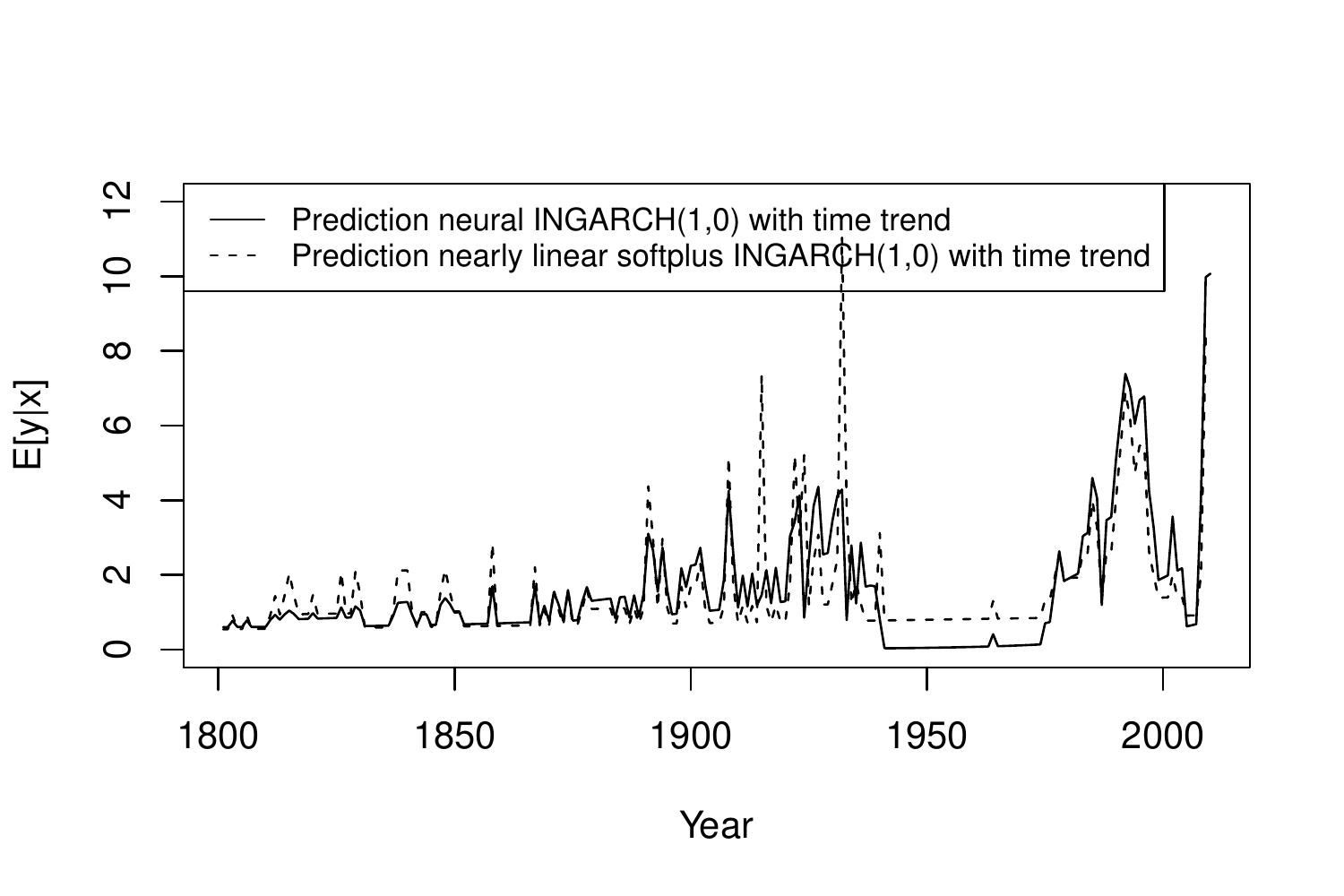} & \includegraphics[trim=0cm 0.5cm 0cm 1cm, clip,scale=0.45]{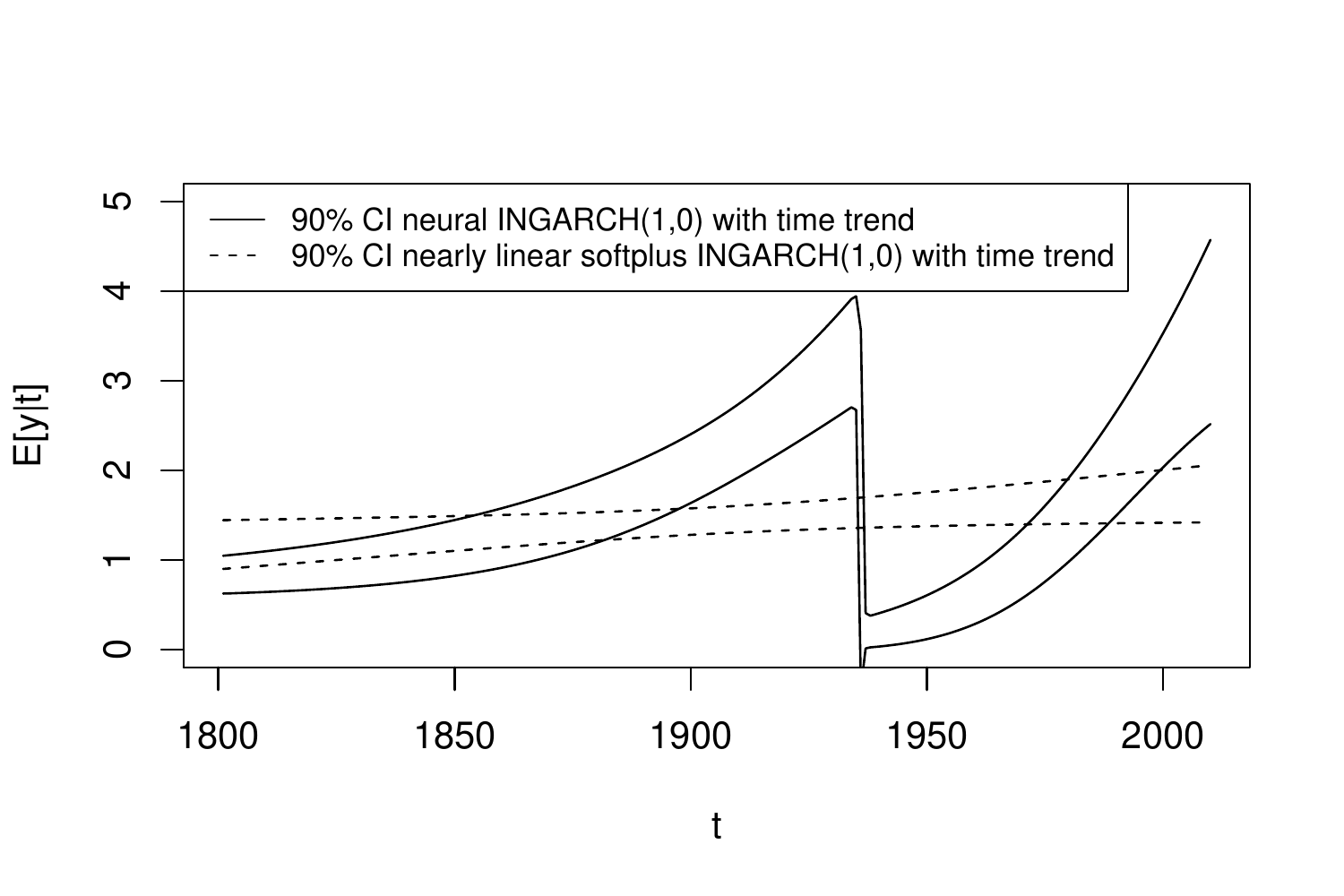} \\
(a)\quad Predicted cond.\ means & (b)\quad Marginal effect of time (at the mean)\\
\includegraphics[trim=0cm 0.5cm 0cm 1cm, clip,scale=0.45]{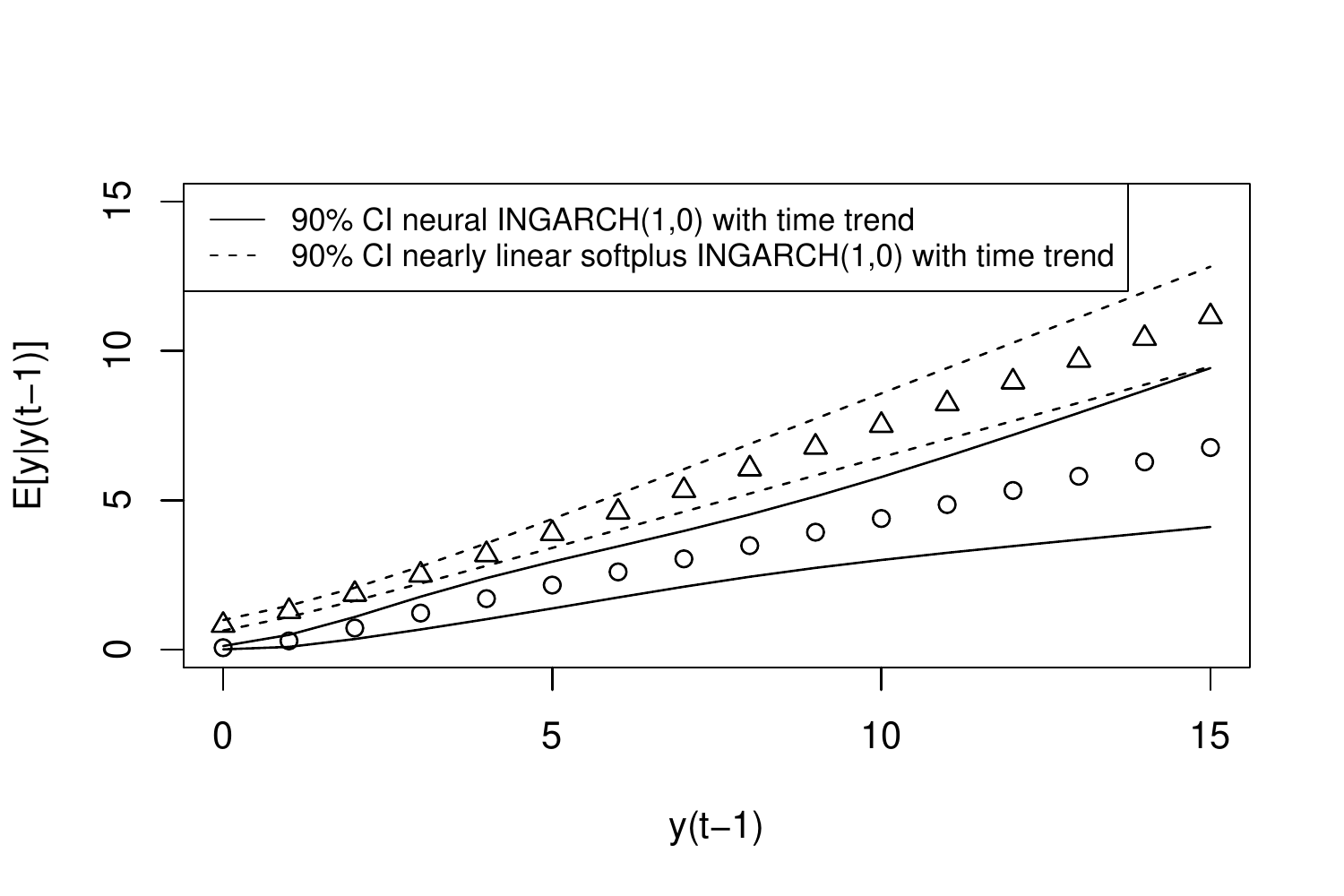} & \includegraphics[trim=0cm 0.5cm 0cm 1cm, clip,scale=0.45]{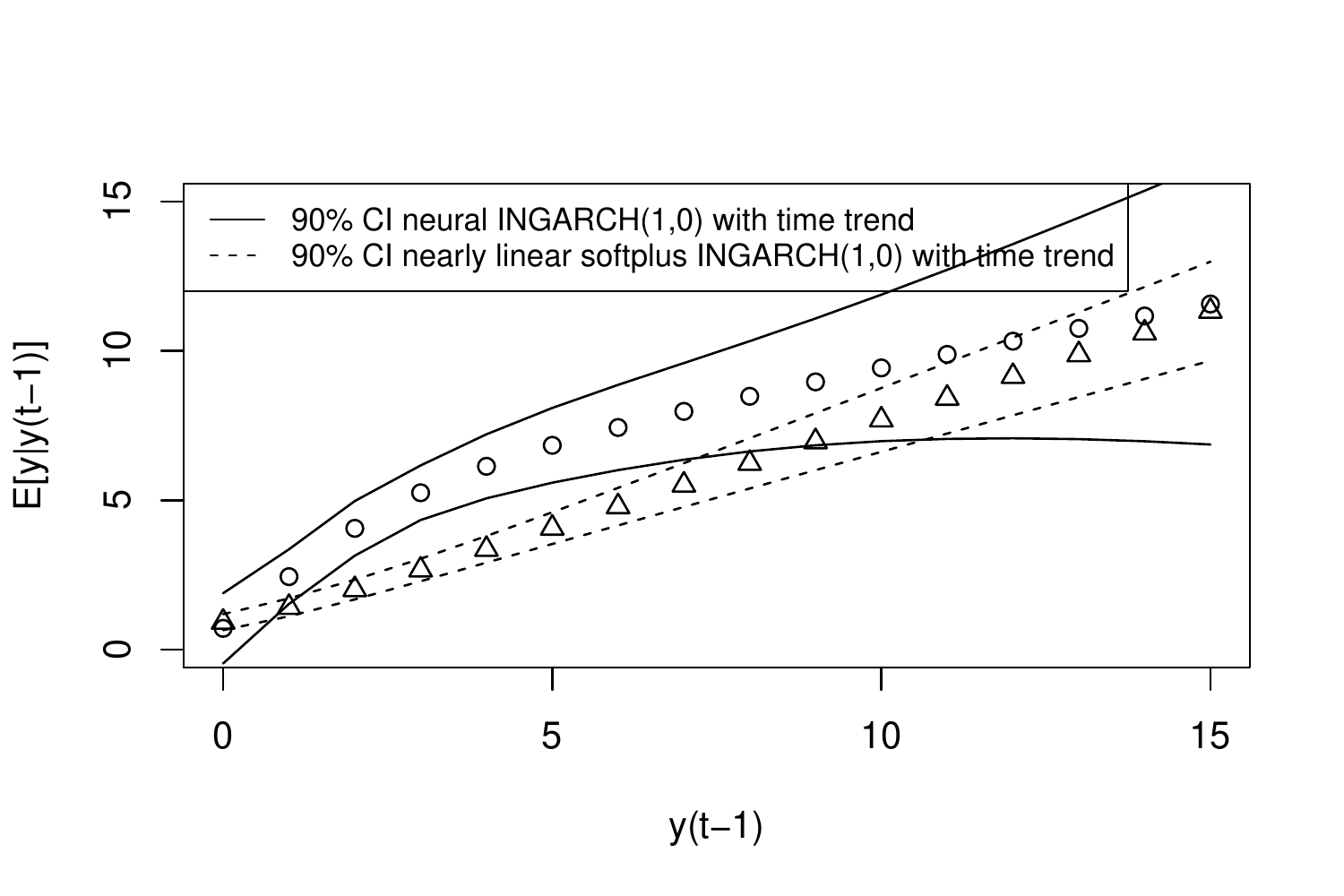} \\
(c)\quad Marginal effect of $y_{t-1}$ at $t=1958$ & (d)\quad Marginal effect of $y_{t-1}$ at $t=2009$\\
\end{tabular}
\caption{Banking crises: Illustrations for the neural INGARCH(1,0) model with time trend}
\label{timetrend}
\end{figure}

In order to investigate in particular the dispersion behavior, the Pearson residuals are analyzed next. Table \ref{pearsontab} (top) shows the mean, variance and the first two ACF values of the empirical Pearson residuals for the four models that have been considered so far. The biggest concern are the variances which are far away from unity, the value expected under a correctly specified model. Fortunately, there are several ways to modify the neural INGARCH model.\\
The first one aims to capture the zero counts during the BW period more accurately. Therefore, a time variable is added as an additional regressor, so the input vector becomes $x=(1,Y_{t-1},t)$. Such a time variable can also be added to the conventional nearly linear softplus INGARCH but it is clear in advance that a nearly linear time trend will not be able to represent the zero counts during the BW period which are only temporary. On the contrary, the universal approximation property of the ANN regression function allows the neural INGARCH model to represent an arbitrary time trend. The number of hidden neurons is increased to $H=3$ which gives the model the required complexity (model selection procedure not shown again). The results are displayed in the middle section of Table \ref{order}. As expected, (only) the neural model benefits significantly from the additional time variable regarding the information loss. For interpretation and comparison, consider Figure \ref{timetrend}. Subfigure \ref{timetrend}~(a) shows the predicted conditional means analogously to Figure \ref{prediction}. The main difference is the BW period where the neural INGARCH model is now able to produce much lower conditional means, as intended.  Figure \ref{timetrend}~(b) shows the marginal effect of time on $E[y_t|x]$ where $y_{t-1}$ is fixed at its mean value (marginal effect at the mean). The ``reset'' of the global banking system in the 1940s is clearly visible. The number of banking crises tends to increase again after the BW period. The BW period can be analyzed more closely by looking at subfigure \ref{timetrend}~(c) which displays the marginal effect of $y_{t-1}$ at time $t=1958$ (during the BW period). The previously discussed probability of staying in the zero-count state is considered again (for that particular year). This time, we have $p_{0|0}^{sp}=0.44$ and $p_{0|0}^{ANN}=0.94$ which confirms the benefit of the time trend in the neural INGARCH model. Finally, subfigure \ref{timetrend}~(d) is analogous to \ref{timetrend}~(c) only that the marginal effect of $y_{t-1}$ is calculate at $t=2009$ and serves to illustrate that the neural model now allows the marginal effect of $y_{t-1}$ to vary across time.\\

The Pearson residuals in Table \ref{pearsontab} (middle) reveal that there is still some overdispersion which the neural INGARCH(1,0) cannot account for, even with a time trend. Therefore, as a final step of the analysis of the banking crises, a generalized Poisson distribution is considered as the conditional distribution. Following \citet{famoye93}, its probability function with parameters $\lambda,\alpha>0$ is defined for non-negative integers $k$ as:
\begin{equation}\label{genpois}
P(X=k)=\left(\frac{\lambda}{1+\alpha\lambda}\right)^k \frac{(1+\alpha k)^{k-1}}{k!}\exp\left(\frac{-\lambda(1+\alpha k)}{1+\alpha \lambda}\right).
\end{equation}

This particular definition of the generalized Poisson distribution is very useful in a regression context because it has $E[X]=\lambda$ and therefore, the regression equation of corresponding INGARCH models is still given by equation \ref{ingarch}. The additional parameter $\alpha$ only affects the variance which is $Var(X)=\lambda(1+\alpha \lambda)^2$. Moreover, the gradient of the log-likelihood function (cf. \citet{famoye93}, eq. 3.3 and 3.4) is simpler than that of the negative binomial distribution which would be a theoretical alternative. The simpler gradient improves the numerical ML estimation, in particular for the neural INGARCH model. The results of the corresponding estimations of the softplus INGARCH(1,0) and the neural INGARCH(1,0) are given at the bottom of Table \ref{order} without further illustrations. It is only noted that $\hat{\alpha}=0.2526$ for the conventional and $\hat{\alpha}=0.1304$ for the neural model, both significant. Finally, a look at the Pearson residuals (Table \ref{pearsontab}, bottom) shows that an acceptable representation of the dispersion structure is now achieved for the neural INGARCH model.

\begin{table}[!h]
  \centering
    \begin{tiny}
    \begin{tabular}{cccccc}
    \toprule
         response&cond. distribution& time trend&order& AIC&BIC\\
    \midrule
	softplus&Poisson&no&(1,0)&713.66&720.36\\
	softplus&Poisson&no&(2,0)&714.20&724.23\\
	softplus&Poisson&no&(1,1)&715.66&725.70\\
	neural&Poisson&no&(1,0)&709.63&729.72\\
	\midrule
	softplus&Poisson&yes&(1,0)&712.39&722.43\\
	neural&Poisson&yes&(1,0)&634.75&674.91\\
	\midrule
	softplus&gen. Poisson&yes&(1,0)&652.26&665.65\\
	neural&gen. Poisson&yes&(1,0)&613.71&657.23\\
	\bottomrule
    \end{tabular}
\end{tiny}
\caption{Banking crises: Information loss for different models}
\label{order}
\end{table}

\begin{table}[!h]
  \centering
  \begin{tiny}
    \begin{tabular}{cccccccc}
    \toprule
         response&cond. distribution& time trend&order& mean($r_t$)&var($r_t$)&ACF(1)($r_t$)&ACF(2)($r_t$)\\
    \midrule
	softplus&Poisson&no&(1,0)&-0.014&2.527&-0.052&-0.025\\
	softplus&Poisson&no&(2,0)&-0.013&2.531&-0.044&-0.032\\
	softplus&Poisson&no&(1,1)&-0.014&2.526&-0.054&-0.024\\
	neural&Poisson&no&(1,0)&0.001&2.620&-0.060&-0.016\\
	\midrule
	softplus&Poisson&yes&(1,0)&-0.010&2.452&-0.052&-0.027\\
	neural&Poisson&yes&(1,0)&0.003&1.585&0.048&-0.025\\
	\midrule
	softplus&gen. Poisson&yes&(1,0)&-0.013&1.490&-0.040&-0.048\\
	neural&gen. Poisson&yes&(1,0)&-0.010&1.099&0.011&-0.044\\
	\bottomrule
    \end{tabular}
\end{tiny}
\caption{Banking crises: Properties of (empirical) Pearson residuals $r_t$ for different models}
\label{pearsontab}
\end{table}

To summarize, it was demonstrated that the neural INGARCH model constitutes a straightforward way to employ ANN regression techniques for the analysis of count time series. From the (conventional) INGARCH perspective, the ANN regression function takes the role of an ``upgraded'' nonlinear response function which includes the conventional model as a degenerate special case. With an appropriate choice of the involved activation functions, the gradient of the neural INGARCH model can be calculated efficiently which allows for fast likelihood maximization. Information criteria can be used to determine the optimal order $(p,q)$ and the optimal complexity of the network $H$. Regarding inference, concepts from non-linear regression such as marginal effects at the mean and numerical Hessian matrices are employed. For the present heavily overdispersed counts, a generalized Poisson distribution with additional variance parameter is useful.\\
A final remark concerns the implications of the results, in particular in relation to those obtained by \citet{dungey20}. One of their main models is a threshold INARMA model (TINARMA) which leads to the identification of two regimes which they label ``systemic'' and ``non-systemic''. Among other aspects, this implies that the number of (countries experiencing) banking crises is mainly explained endogenously. On the contrary, the (non-linear) time trend suggested in this paper offers an exogenous explanation and enables the model to represent the zero counts during the BW period very accurately.

\section{Modeling bounded counts}\label{binarch}
As a second example, bounded counts (with upper limit $n$) are considered. Then, the binomial distribution is employed as the conditional distribution instead of the Poisson distribution. Because of the parametrization of the binomial distribution, the INGARCH regression equation now refers to the (conditional) success probability instead of the (conditional) mean. The binomial INGARCH model is defined by
\begin{equation}\label{binomial}
Y_t \sim Bin(n, P_t) \mbox{ with } P_t=f(Y_{t-1},\dots,Y_{t-p},P_{t-1},\dots,P_{t-q}).
\end{equation}
The main consequence for the ANN architecture is that the activation function $g_{1}$ has to yield a probability and therefore, should map to the interval $[0,1]$. An obvious choice is the logistic function which is generally employed as the activation function for the hidden layer $g_0$ in this paper, so $g_0(x)=g_1(x)=\frac{1}{1+\exp(-x)}$ for the neural binomial INGARCH model. As explained previously, the function $g_1$ takes the role of the response function in conventional INGARCH models, meaning here that the conventional binomial INGARCH model with logit link (i.e. logistic response) function \citep[cf.][]{chen20} can be interpreted as the corresponding degenerate version of the neural binomial INGARCH model. An alternative choice would be the soft-clipping activation function. This function constitutes an almost linear response function and the idea is similar to the softplus response function for unbounded counts. A corresponding soft-clipping binomial INGARCH model is investigated by \citet{weissjahn21}.\\

\begin{figure}[!ht]
\center\footnotesize
\begin{tabular}{c@{\qquad}c}
\multicolumn{2}{c}{\includegraphics[trim=0cm 0cm 0cm 1.5cm, clip,scale=0.65]{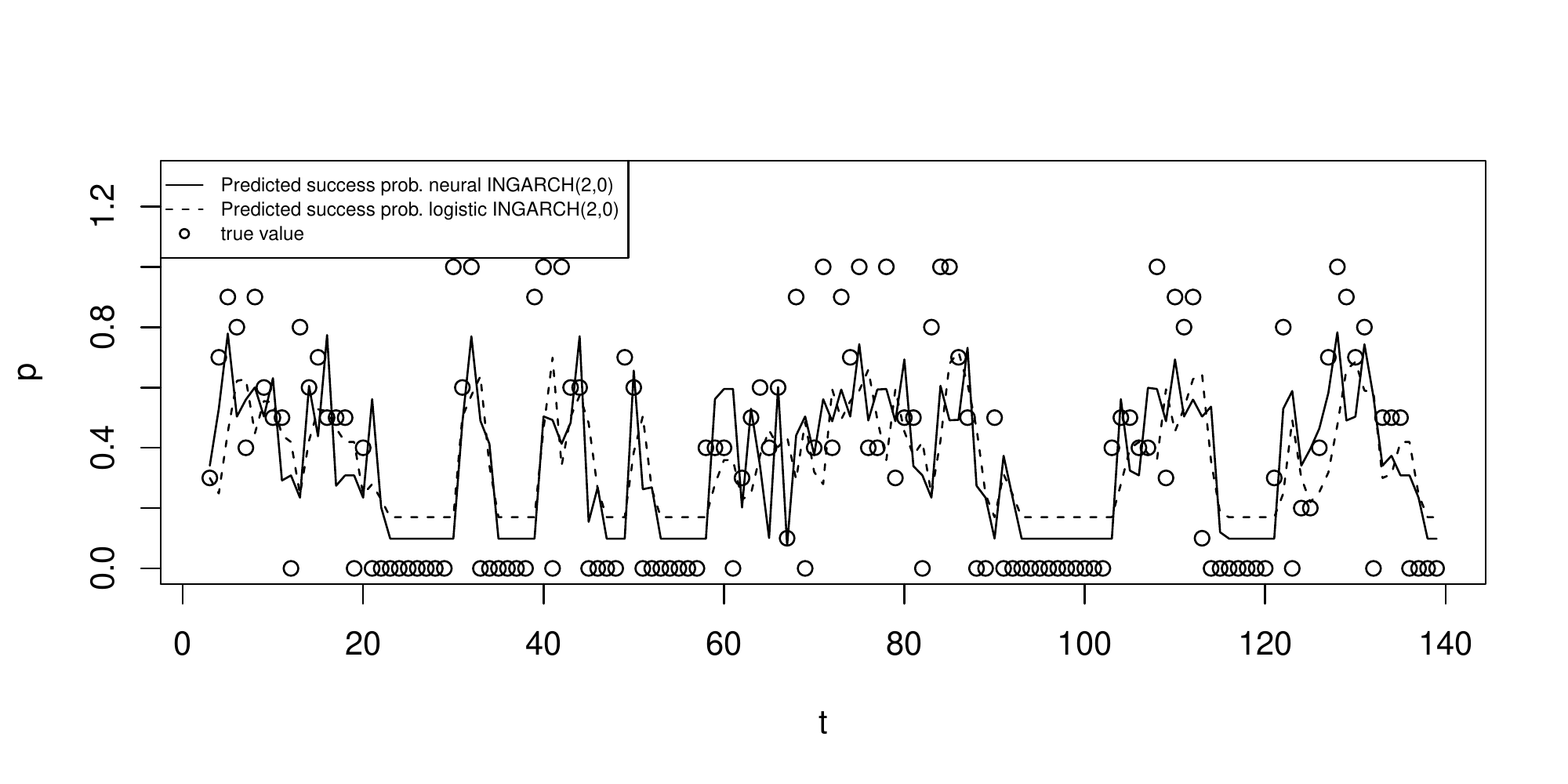}}\\
\multicolumn{2}{c}{(a) Predicted cond.\ success probabilities}\\
\includegraphics[trim=0cm 0cm 0cm 0cm, clip,scale=0.45]{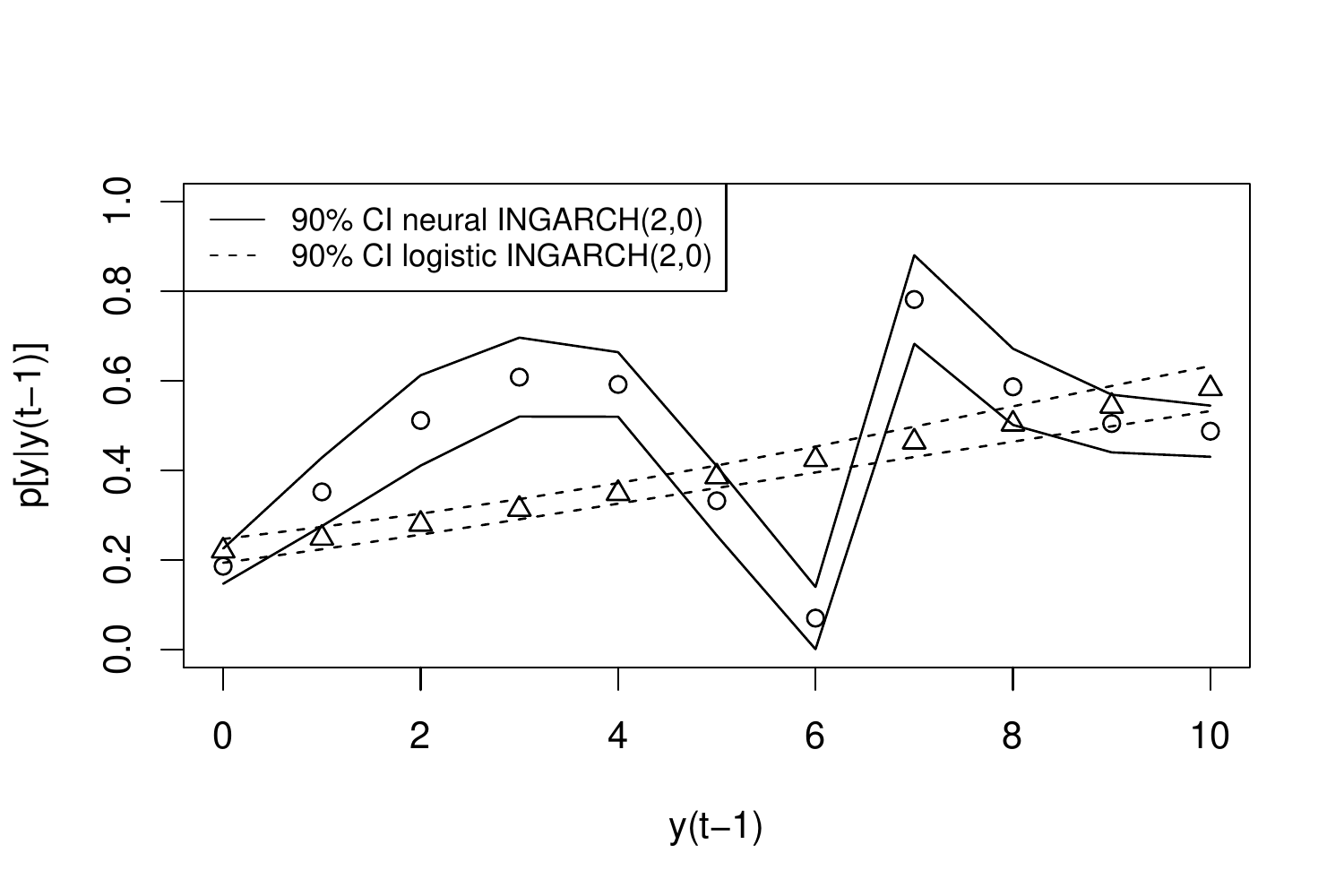} & \includegraphics[trim=0cm 0cm 0cm 0cm, clip,scale=0.45]{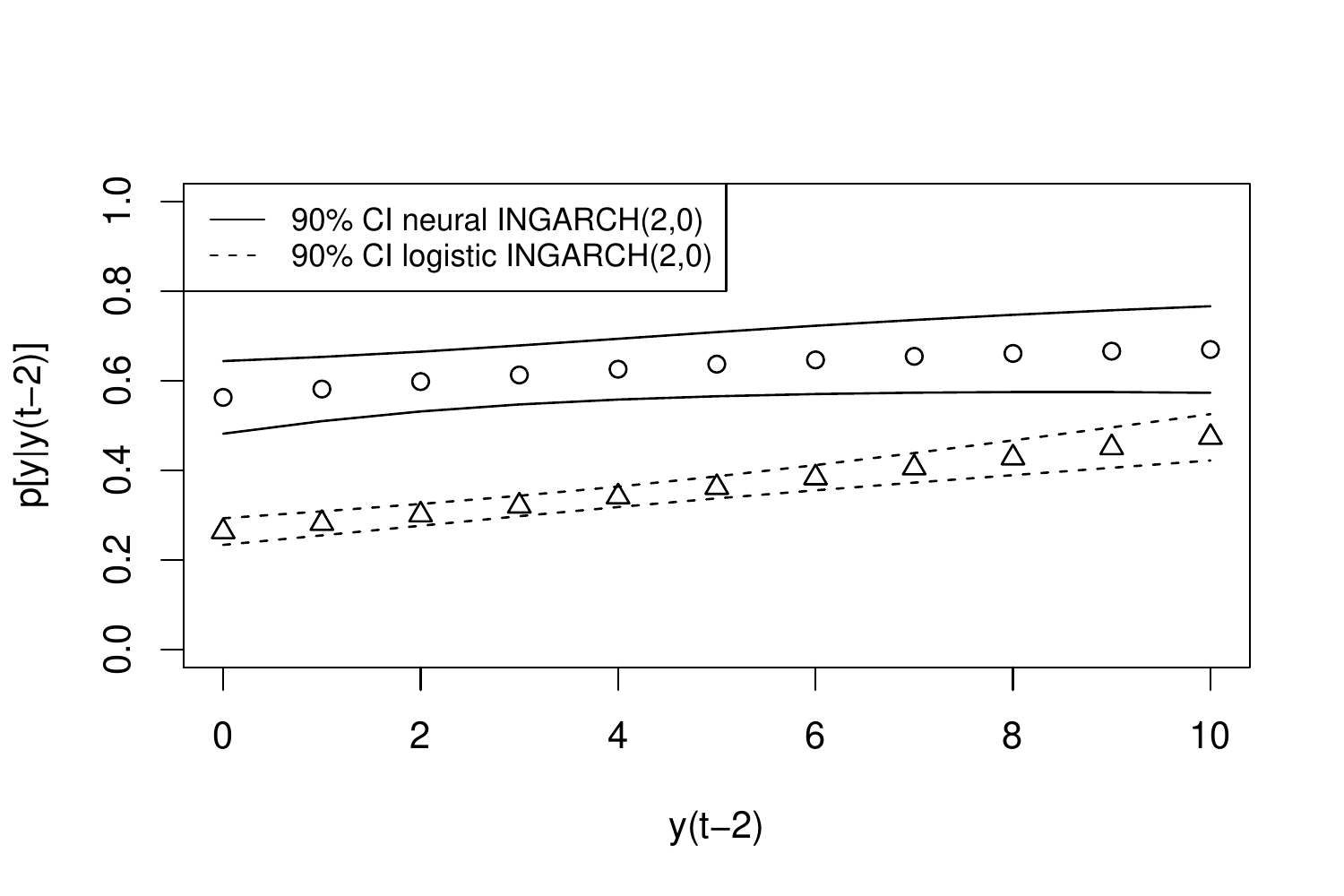} \\
(c)\quad Marginal effect of $y_{t-1}$& (d)\quad Marginal effect of $y_{t-2}$\\
\end{tabular}
\caption{MPC votes: Predictions and marginal effects}
\label{mpc}
\end{figure}

As the data example, the monetary policy decisions of the Monetary Policy Council (MPC) of the National Bank of Poland are considered. Precisely, the data are the number of MPC members voting in favor of an adjustment of the interest rate at the monthly meetings between 2002 and 2013 (139 months). There are $n=10$ members so we have a monthly time series of bounded counts. The data were originally analyzed by \citet{sirchenko13,sirchenko20} and then also by \citet{moeller18}.\\
The procedure to determine an appropriate binomial neural INGARCH model is the same as for the previously considered Poisson neural INGARCH model for unbounded counts. First, the INGARCH model order $(p,q)$ is selected based on the conventional counterpart which is the binomial INGARCH model with logistic response function \citep[cf.][]{chen20}. Table \ref{orderbino} shows the information loss for different model orders ($p+q\leq 3$ considered). The BIC selects $(p,q)=(2,0)$ by some margin and the AIC does not provide a clear decision between $(p,q)=(2,0)$ and $(p,q)=(3,0)$, so the BIC-optimal order is chosen.  Regarding the network complexity $H$, the rule of thumb yields $H\leq\frac{0.1 T}{J+1}=\frac{13.7}{4}=3.425$ and, therefore, $H=3$ is selected. The corresponding AIC and BIC values are displayed in Table \ref{orderbino} and they imply that updating the logistic model to a neural model is sensible in terms of the information loss. Figure \ref{mpc} helps explaining possible reasons. First note that the observed counts are characterized by multiple zero runs of medium length on the one hand and a relatively large share of high values on the other hand. The predicted conditional success probabilities (Figure \ref{mpc}~(a)) from both models mostly stay at a medium level which is ``the best they can do''. Similar to the previous example of bounded counts, the neural model achieves a higher probability of staying in the zero-count state (assuming $y_{t-1}=y_{t-2}=0$). The calculations yields $p_{0|0,0}^{logi}=0.16$ and $p_{0|0,0}^{ANN}=0.35$. The marginal effect of $y_{t-1}$ in the neural model resembles a polynomial but is in general agreement with the positive marginal effect from the logistic model. From subfigure (c), it could be suspected that the effect of $y_{t-2}$ is insignificant and that allowing for a nonlinear effect of $y_{t-1}$ eliminates the need to consider $y_{t-2}$.

The analysis of the Pearson residuals (Table \ref{pearsontabbino}, top) shows that none of the models captures the dispersion behavior of the monetary policy votes in a satisfactory way. Still, the neural model produces the most adequate fit which is in line with the lower information loss (Table \ref{orderbino}). An extended model with a time trend was also tested but, other than in the previous data example, the goodness-of-fit could not be improved further due to the nature of the zero runs. \citet{moeller18} consider various zero-inflated models for the same time series.  In this paper, the example primarily serves to motivate the proposed novel neural INGARCH framework, so only the simplest option to implement zero-inflation will be discussed. The conditional binomial distribution is then replaced by the following zero-inflated binomial (ZIB) distribution with parameters $\omega, n$ and $p$:

\begin{equation}\label{zib}
P(X=k)=\mathbbm{1}_{[k=0]}\omega  + (1-\omega)  P_{n,p}^{bino}(X=k).
\end{equation}

\begin{figure}[!h]
\center\normalsize
\includegraphics[width=\textwidth, trim=0cm 0cm 0cm 1.5cm, clip]{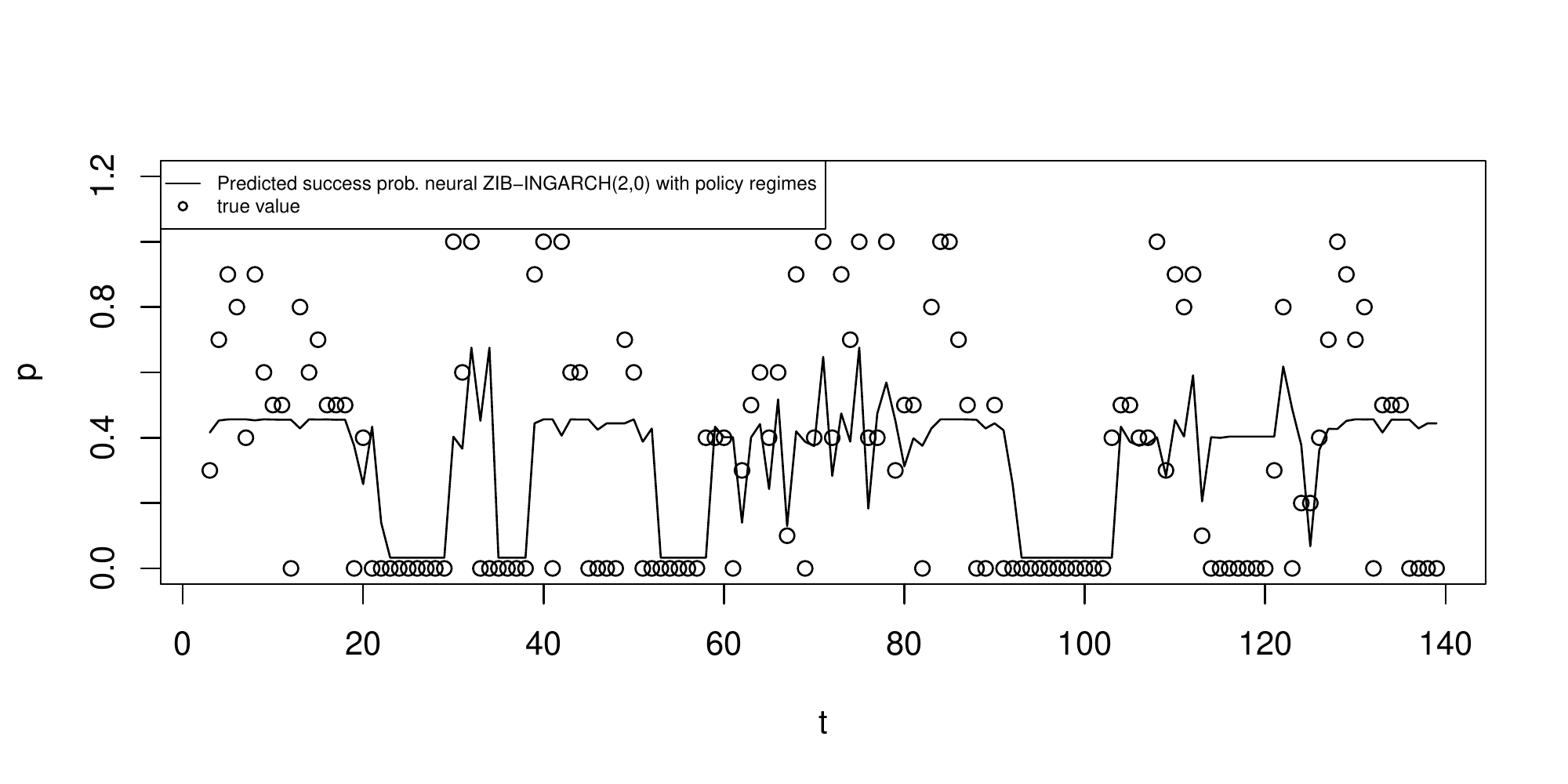}
\caption{MPC votes: Predicted cond.\ success probability by the neural ZIB-INGARCH(2,0) model with policy regime variable}
\label{predregime}
\end{figure}

It can be useful to reparametrize with $\pi=(1-\omega)p$ because the expected value is $E[X]=n\pi$. The variance is $Var(X)=n\pi(1-\pi)+n(n-1)\frac{\omega}{1-\omega}\pi^2$ \citep[cf.][eq.\ 2.4]{moeller18}, so corresponding INGARCH models feature (conditional) extra-binomial variation (for $\omega>0$). Estimating the two ZIB-INGARCH(2,0) models, a massively reduced information loss is obtained (Table \ref{orderbino}, middle). Therefore, it is not surprising that the zero-inflation parameter is quite large, $\hat{\omega}=0.43$ for both models. However, considering Table \ref{pearsontabbino} (middle), it is found that the better representation of the dispersion structure comes at the cost of serial correlation within the Pearson residuals. Such a trade-off is also observed by \citet{moeller18} where the two best models in terms of the variance of the Pearson residuals (closest to 1) feature high levels of serial correlation \citep[][Table 6]{moeller18}. Referring to \citet{sirchenko20} a possible solution is to look more closely at the nature of the zeros. The meaning of the zero count is that all MPC members vote for the interest rate to remain unchanged. He distinguishes three different situations when this can occur: in periods of policy tightening; in periods of maintaining between rate reversals; and in periods of policy easing \citep[][p.\ 1]{sirchenko20}. In the neural INGARCH model, these regimes can be represented by just a single additional variable with value $-1$ (easing), $0$ (maintaining) or $1$ (tightening). This result in a model with different regimes for each value of the policy variable. In particular, the probability of staying in the zero-count state now depends on the policy regime. The estimation yields $p_{0|0,0,-1}\approx p_{0|0,0,1}=0.32$ and $p_{0|0,0,0}=0.74$, so a distinction between at least two regimes should be made. Figure \ref{predregime} displays the corresponding predicted cond.\ success probabilities and indeed, the zero runs are now represented much better. This is confirmed by the reduced information loss (Table \ref{orderbino}, bottom). In fact, the model outperforms the best model discussed in \citet[][Table 5]{moeller18} in terms of AIC and BIC. Considering the Pearson residuals, the extended model offers a compromise between the variance and the autocorrelation (Table \ref{pearsontabbino}, bottom). The (two-sided) 95\% critical value for the ACF is $\frac{1.96}{\sqrt{137}}=0.167$, so the autocorrelation is not (highly) significant.

\begin{table}[!h]
  \centering
    \begin{tiny}
    \begin{tabular}{cccccc}
    \toprule
         response&cond. distribution& policy regimes&order& AIC&BIC\\
    \midrule
	logistic&binomial&no&(1,0)&1029.1&1034.9\\
	logistic&binomial&no&(2,0)&989.0&997.8\\
	logistic&binomial&no&(3,0)&988.3&1000.0\\
	logistic&binomial&no&(1,1)&1009.8&1018.6\\
	logistic&binomial&no&(2,1)&990.9&1002.6\\
	logistic&binomial&no&(1,2)&989.6&1001.3\\
	neural&binomial&no&(2,0)&868.0&903.1\\
	\midrule
	logistic&ZIB&no&(2,0)&610.0&621.6\\
	neural&ZIB&no&(2,0)&581.9&619.9\\
	\midrule
	neural&ZIB&yes&(2,0)&532.0&578.7\\
	\bottomrule
    \end{tabular}
\end{tiny}
\caption{MPC votes: Information loss for different models}
\label{orderbino}
\end{table}

\begin{table}[!h]
  \centering
  \begin{tiny}
    \begin{tabular}{cccccccc}
    \toprule
         response&cond. distribution& policy regimes&order& mean($r_t$)&var($r_t$)&ACF(1)($r_t$)&ACF(2)($r_t$)\\
    \midrule
	logistic&binomial&no&(1,0)&-0.022&5.256&-0.039&0.188\\
	logistic&binomial&no&(2,0)&-0.031&5.105&0.037&0.036\\
	logistic&binomial&no&(3,0)&-0.031&5.145&0.034&0.033\\
	logistic&binomial&no&(1,1)&-0.043&5.131&0.023&0.090\\
	logistic&binomial&no&(2,1)&-0.030&5.111&0.033&0.032\\
	logistic&binomial&no&(1,2)&-0.062&5.120&0.023&0.036\\
	neural&binomial&no&(2,0)&0.006&4.853&0.028&0.002\\
	\midrule
	logistic&ZIB&no&(2,0)&0.004&1.197&0.389&0.323\\
	neural&ZIB&no&(2,0)&0.029&1.141&0.381&0.307\\
		\midrule
	neural&ZIB&yes&(2,0)&0.017&1.470&0.193&0.151\\
	\bottomrule
    \end{tabular}
\end{tiny}
\caption{MPC votes: Properties of (empirical) Pearson residuals $r_t$ for different models}
\label{pearsontabbino}
\end{table}

Concluding the MPC votes data example, it was shown that the neural INGARCH model, with a suitable choice of the involved activation functions, can also be employed for time series of bounded counts. Both layers in the neural model are then activated by the logistic function as the regression equation refers to the conditional success probability. The higher likelihood of the neural model compared with the corresponding conventional INGARCH model generally justifies the larger number of parameters. The zero-inflation can be accounted for by using a conditional ZIB distribution instead of the binomial distribution. Finally, as an extension beyond the pure INGARCH model, a policy regime variable helps explaining the zero runs in times of stable interest rates.

\section{Conclusion and outlook}
The paper introduced a novel class of INGARCH models where the response function corresponds to an artificial neural network, more precisely, a single hidden layer feedforward network. Due to the universal approximation property, neural INGARCH models are very flexible regarding possible nonlinear influences of the considered regressors and interaction effects between them. Conventional INGARCH models can be interpreted as ``degenerate'' instances of the corresponding neural INGARCH models. This implies that the activation function of the output neuron in the SLFN takes the same role as the response function in a conventional INGARCH model. Conversely, a neural INGARCH model can be interpreted as an ``upgraded'' version of a corresponding  conventional model. Two data examples for unbounded and bounded counts demonstrate the empirical relevance as the neural INGARCH models are preferred over their conventional counterparts in terms of the information loss. Additional explanatory variables such as a time variable or a regime variable can be easily incorporated. Apart from the computationally demanding numerical likelihood maximization, the main limitation of the neural INGARCH model is the relatively high number of parameters which makes it difficult to apply it to short time series with $T<100$.\\
Regarding future research, the possibilities to extend the use of ANN are manifold. First, the framework can obviously be generalized to ANN regression functions with more than one hidden layer (deep networks). The associated research question could then be whether or in which situation more layers with less neurons are preferred over one layer with more neurons. Furthermore, the nonlinear modeling capability of the neural INGARCH model could be utilized for multivariate analysis. For example, multiple output neurons could be considered to obtain a type of nonlinear VAR model. In case of spatial data, spatio-temporal INGARCH model could provide more accurate estimation of how disturbances spread over space and time compared to linear models.

\end{document}